\documentclass[aps,twocolumn]{revtex4}
\usepackage[dvips]{graphicx}
\usepackage{psfrag}

\usepackage{dcolumn}
\usepackage{bm}
\begin{document}

\title{Time-dependent theory of non-linear response and current fluctuations}
\author{I. Safi}

\affiliation{Laboratoire de Physique des Solides, Universit\'e
Paris-Sud, 91405 Orsay, France}

\date{\today}

\begin{abstract}
A general non-linear response theory is derived for an arbitrary time-dependent Hamiltonian, not necessarily obeying time-reversal symmetry. This allows us to obtain a greatly generalized Kubo type formula. Applied to a mesoscopic system with any type of interactions, and coupled to multiple probes and gates with arbitrarily time-dependent voltages, we derive current-conserving differential conductance and current fluctuation matrices obeying a generalized Fluctuation-Dissipation Theorem. This relation provides a common explanation for asymmetries of the excess noise in several non-linear mesoscopic systems, as well as of its surprising negative sign.

\end{abstract}

\maketitle

The linear response theory is a cornerstone of the quantum theory. It allows deriving extremely useful formulas such as the Kubo formula or the Fluctuation- Dissipation Theorem (FDT) which are extensively used in all fields. It has been somewhat a common belief that the validity of the Kubo formula is limited essentially to the linear regime, motivating other alternatives in correlated systems \cite{negf}. In the present work we show that, in fact, it is possible to greatly generalize the Kubo formula within a non-linear response theory capable of addressing non-equilibrium situations and to include interactions as well as any (possibly time-dependent) Hamiltonian. Our work goes beyond and is more general than previous extensions performed in the sixties \cite{peterson} and more recently in the stationary regime\cite{gavish}: these works were not adapted to a mesoscopic context and to time-dependent Hamiltonian or/and voltages, and even under these limitations we are able to yield a much more compact proof.

Even though our theory is not restricted to condensed matter theory, we choose to apply it to develop a novel transport formalism for arbitrary mesoscopic systems connected to two or many probes with any time dependence of their electrochemical potentials. Such multi-probe geometries are of great interest for e.g. revealing statistics and entanglement, as in Hanbury-Brown Twiss setups or Mach-Zhender interferometers. For these systems, we express the differential out-of-equilibrium conductance matrix in a microscopic way, ensuring the current conservation and gauge invariance. We thereby provide a convenient general framework to describe finite frequency and/or time-dependent behavior of these multi-probe systems which was lacking up to now. We also solve subtleties which have been a subject of debates since the development of the scattering approach \cite{buttiker_revues,buttiker}. Our formalism is an important achievement not only for time-dependent voltages or/and Hamiltonian, but is already crucial for the stationary regime. It offers a promising alternative to other approaches \cite{buttiker_revues,buttiker,multi_photo,polianski_buttiker,negf,multi_negf}, which even though successful and extended to nonlinearities and/or AC voltages \cite{note3}, are not suited to deal in a systematic way with a majority of the strongly correlated systems \cite{note1}.

Since we provide for any time-dependence of voltages, we can consider not only AC voltages \cite{gabelli_07} to study photo-assisted transport \cite{lesovik_photo,exp_photo,crepieux_photo}, but also injection of electrons on demand \cite{feve_07,safi,keeling_demand}, classical sources of noise, pumping, or mixing setups where the potentials in the reservoirs or gates have different periods, etc. We can also consider spontaneous generation under a DC bias, such as finite frequency (FF) noise \cite{buttiker_revues}, or combine both, for instance by applying time-dependent voltages and consider FF current fluctuations, which in this situation depend on two frequencies \cite{note_reulet} and form a matrix containing both auto-correlations and cross-correlations. Using our out-of-equilibrium and time-dependent Kubo formula, we show that such a matrix obeys a general time-dependent out-of-equilibrium FDT. Then we discuss its important consequences in the limit of stationary Hamiltonian and voltages. A first application of the present formalism has been done in quantum wires and carbon nanotubes \cite{ines_ff_noise}.



 For generality, we consider a system with an arbitrary time-dependent Hamiltonian $\mathcal{H}(t)$, which includes any interactions or disorder, and does not necessarily obey time-reversal symmetry. $\mathcal{H}(t)$ depends in linear/non-linear and in local/non-local way on a set of time-dependent parameters generically denoted by $X(t')$. We express the functional derivative of the average of any operator $O$ at time $t$ with respect to $X(t')$. For this purpose, the Hamiltonian is split into the part which does not depend on $X$, denoted by $\mathcal{H}_0(t)$, and another which depends on $X$: $\mathcal{H}(t)=\mathcal{H}_0(t)+\mathcal{H}_1(t,X)$. One switches to the interaction picture where $\mathcal{H}_1$ is viewed as the interaction Hamiltonian. Then $O^{int}(t)=U_0(-\infty,t)O(t)U_0(t,-\infty)$ where $i\hbar \partial_t U_0(t,-\infty)=\mathcal{H}_0(t)U_0(t,-\infty)$. Even though not necessary, we prefer to exploit the Keldysh formulation to make our argument more compact. The Keldysh time contour has two branches labeled by $\eta=\pm$, going from $-\infty$ to $\infty$ on the upper one and inversely on the lower one. $T_K$ is the Keldysh ordering operators which makes time (anti-time) ordering on the upper (lower) contour, while operators labeled by $-$ are on the left to these labeled by +:  $\langle T_K A^+(t)B^-(t')\rangle=\langle B(t')A(t)\rangle$. $O$ can be labeled by $\eta=+$ or $-$ for its average expression, of which the functional derivative with respect to $X(t')$ is: \begin{eqnarray}
&&\frac{\delta \langle O(t)\rangle}{\delta X(t')}=\frac{\delta}{\delta X(t')}\langle T_K O^{+}(t)e^{-\frac{i}{\hbar}\int_{-\infty}^{\infty}{\small\sum_{\eta}}\eta \mathcal{H}_1^{\eta}(s)ds}\rangle\nonumber\\
&&=-\frac i{\hbar}\langle T_K \int dt"\sum_{\eta"}\eta" O^{+}(t)\frac{\delta \mathcal{H}^{\eta"}(t")}{\delta X(t')}e^{-\frac{i}{\hbar}\int{\small\sum_{\eta}}\eta \mathcal{H}_1^{\eta}(s)ds}\rangle.\nonumber
\end{eqnarray}
 Using $\sum_{\eta"}{\eta"} \langle T_K A^+(t)B^{\eta"}(t")\rangle\!\! =\!\!\theta(t-t")\langle [A(t),B(t")]\rangle$, we obtain the central result of this paper:
\begin{eqnarray}\label{general_response}
\frac{\delta \langle O(t)\rangle}{\delta X(t')}&=&\frac{-i}{\hbar}\int dt"\theta(t-t")\left\langle \left[O(t),\frac{\delta \mathcal{H}(t")}{\delta X(t')}\right]\right\rangle.
\end{eqnarray}
Thus the functional derivative of the average of an operator $O$ can be expressed in terms of its commutator with the generalized force, defined by the functional differential of ${\mathcal{H}}(t)$. The average is taken in the presence of ${\mathcal{H}}(t)$ and with an initial density matrix which has not to be thermal.  We will express higher order differentials, and allow for $O$ to depend on $X$ separately \cite{ines_next}.  \\
\begin{figure}[b]
\psfrag{Vg}[][][1.]{$V_0$}
\psfrag{I}{$I_n(t)$}\psfrag{V'}{$V_{n'}(t')$}
\psfrag{V''}{$V_{n''}(t'')$}\psfrag{H}{${\cal H}(t,X)$}
\centering\includegraphics[width=6.cm]{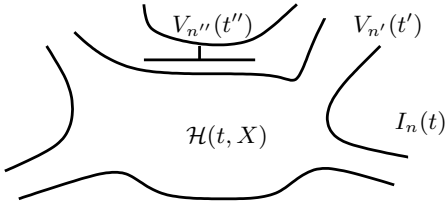} \label{figure1}
\caption{\small A mesoscopic system coupled to many terminals including gates with arbitrary time-dependence of their electrochemical potentials. The time-dependent Hamiltonian $\mathcal H$ describes any interactions or disorder and can depend on other parameters $X(t')$ in a non-local and nonlinear way. Differential of the current average $\langle I_n\rangle$ at terminal $n$ either with respect to $V_{n'}(t')$ or to $X(t')$ can be expressed through a generalized response formula keeping all $V_n$ and $X$ finite. }
\end{figure}
Now we apply this formula to a mesoscopic system connected to N terminals with electro-chemical potential $\mu_{n}(t)=eV_{n}(t)$ and a total charge operator $Q_n$, the current operator $I_n$ for each $n=1,.., N$ being $I_n=-\partial_tQ_n$. The system is described by $\mathcal{H}_0(t)$, which can for instance include time-dependent scatterers, to which we add the coupling to terminals \cite{buttiker,note1}:
 \begin{equation}\label{coupling}{\mathcal H}_{1}(t)=\sum_{n=1}^{M}Q_{n}V_{n}(t).\end{equation}
 It is possible to justify this coupling \cite{ines_next} by generalizing the formalism developed in one-dimensional wires \cite{safi,ines_ff_noise}.
  We include the gates in the $N$ terminals. We do not use the currently adopted coupling in terms of capacitances. Actually, this would be too specific as the latter could acquire
 a frequency dependence as well \cite{ines_next}. 
It is now possible to express in a microscopic way the differential out-of-equilibrium conductance without any constraint on the Hamiltonian, the initial density matrix.  We use Eq.(\ref{general_response}) where $X(t)$ is replaced by $V_{n}(t)$ on which the interaction Hamiltonian in Eq.(\ref{coupling}) depends now in a local and linear way. Denoting by $G_{nn'}(t,t')=\delta\langle I_n(t)\rangle/\delta V_{n'}(t')$ for $n,n'=1..N$, we get the crucial result:
\begin{equation}\label{gmatrix}\hbar G_{nn'}(t,t')=\theta(t-t')\left\langle\left[I_{n}(t),Q_{n'}(t')\right]\right\rangle,
\end{equation} where the average is calculated still \textit{in the presence} of the time-dependent Hamiltonian $\mathcal{H}_0(t)+\mathcal{H}_1(t)$. Let's denote by
$
F(\omega,\omega')=\int \int dt dt' e^{i\omega t-i\omega't'} F(t,t'),$ the double-Fourier transform of a function $F$. Then
$G_{nn'}(\omega,\omega')$ is the variation of $\langle I_n(\omega)\rangle$ when an infinitesimal modulation at $\omega'$ is added to $V_{n'}(t')\rightarrow V_{n'}(t')+ v_{n'}(\omega')e^{i\omega't'}$, keeping $V_{n'}(t')$ finite.

  In particular, Eq.(\ref{gmatrix}) should apply to interacting quantum dots weakly coupled to many reservoirs. In this case, it was claimed that the Kubo formula was not appropriate, contrary to what we have shown here, and the so-called non-equilibrium Green's function formalism was proposed as an alternative \cite{negf}. In order to ensure the current conservation and gauge invariance which it missed, it has been adapted to a multi probe geometry where the values of the currents were modified \cite{multi_negf}. We think that this procedure lacks transparency and generality.
   
   Here, by carrying a total charge opposite to that on the system, the gates ensure the conservation law:
$
\sum_{n=1}^N Q_n=0.
$
This guarantees gauge invariance automatically: a translation of all potentials $V_n(t)$ by the same function $V(t)$ has no effect on $\mathcal{H}_1(t)$ in Eq.(\ref{coupling}). Using Eq.(\ref{gmatrix}), we get simultaneously the two constraints $
\sum_{n}G_{nn'}(\omega,\omega')=\sum_{n'}G_{nn'}(\omega,\omega')=0,$ the second one corresponds to gauge invariance. Thus the latter is not required independently of the current conservation, contrary to previous works in the scattering approach.
We emphasize that there is not necessarily time-reversal symmetry, thus no symmetry of the matrix $\bf G$. It is not hermitian neither. One can show that : $G_{nn'}^{*}(\omega,\omega')=G_{nn'}(-\omega,-\omega')$, where the star stands for the complex conjugate.

Now we consider the non-symmetrized current fluctuations matrix whose elements are given by:
\begin{equation}\label{noise}
S_{nn'}(t,t')=\left\langle I_{n'}(t')I_{n}(t)\right\rangle-\left\langle I_{n'}(t')\right\rangle\left\langle I_{n}(t)\right\rangle.
\end{equation}
 Let's consider:
\begin{equation}\label{sa}\mathbf{S^{\pm}}(t,t')=\mathbf{S}(t,t')\pm{\mathbf S^ T}(t',t),
\end{equation}
the symmetric and the antisymmetric parts of the current fluctuation matrix, where the subscript $T$ refers to the transpose. Notice that one has  $S_{n'n}(t',t)=S_{nn'}^*(t,t')$. 
One can show easily that $\hbar\partial _{t'}G_{nn'}(t,t')-\hbar\partial _t G_{n'n}(t',t)=S^-_{nn'}(t,t')$, which,
 once Fourier transformed, gives (see Eqs.(\ref{noise},\ref{gmatrix})):
\begin{eqnarray}\label{compact}
\mathbf{S^-}(\omega,\omega')&=& S_{nn'}(\omega,\omega')-S_{n'n}(-\omega',-\omega)\nonumber\\&&=-\hbar\omega' {\bf{G}}(\omega,\omega')-\hbar\omega {\mathbf G^{\dagger}}(\omega',\omega).
\end{eqnarray}
  This is a novel FDT which extends to out-of-equilibrium, time-dependent Hamiltonian and voltages, and with or without time-reversal symmetry. It allows to relate both symmetrized and non-symmetrized fluctuations (see Eq.(\ref{sa})), as one can inject this expression into the r. h. s. of: \begin{equation}
 \label{symnon}
 2{\bf S}={\bf S^+}+{\bf S ^-}.
 \end{equation}
One can show that $S_{n'n}(\omega',\omega)=S_{nn'}^{*}(\omega,\omega')$. Let's specify to $\omega =\omega'$, which amounts to integrate out over $t-t'$ to get the DC component of the fluctuations in the presence of any set of $V_n(t)$. Eq.(\ref{compact}) reduces to :
\begin{equation}\label{asymmetry_one}
\mathbf{S^-}(\omega,\omega)=-2\hbar\omega {\mathbf G^{h}}(\omega,\omega),
\end{equation} where the Hermitian part of the matrix $\bf G$ is given by:
\begin{equation}2{\mathbf G^h}={\mathbf G}+{\mathbf G^{\dagger}}.\label{hermitian_part}
\end{equation}
We note that  ${\mathbf S}(\omega,\omega)$ and hence $\mathbf S^{\pm}(\omega,\omega)$ become hermitian, which is not the case for $\mathbf{G}(\omega,\omega)$. Only the diagonal elements of these four matrices are real.

Let us now comment briefly on the case of a periodic potential. Allowing for generality a different frequency $\Omega_{n}$ in each terminal $n$ as in mixing setups, one requires  that at least for one terminal $n$ one has $\omega-\omega'=l\Omega_{n}$ with $l$ an integer. But both $\omega$ and $\omega'$ can be different from a multiple of all $\Omega_{n}$ \cite{note_modulation}.

We will focus in the following on both time-independent Hamiltonian and potentials in the reservoirs \cite{ines_cond}. Then invariance by time translation is restored and one requires $\omega'=\omega$ in Eqs.(\ref{gmatrix},\ref{compact}). Let's keep similar notations for $\bf{F}=\bf{S},\bf{G}$ but stress their dependence on the voltage vector ${\bf V}=(V_1,V_2..V_N)$: $$\mathbf{F}(\omega',\omega)=\delta(\omega'-\omega){\bf\mathbf  F_V}(\omega).$$ We also introduce the "excess AC differential conductance" and excess FF noise matrices for later use by:
\begin{equation}\label{excess}
{\bf \Delta F_V}(\omega)={\bf\mathbf  F_V}(\omega)-{\bf\mathbf  F_{V=0}}(\omega).
\end{equation}
As noticed above for the case $\omega=\omega'$, the matrices $\mathbf S_V,S_V^{\pm}$ become hermitian while the matrix $\mathbf G_V$ is not. The asymmetry of the FF fluctuation matrix is analogous to Eq.(\ref{asymmetry_one}) with a unique frequency (see Eq.(\ref{hermitian_part})):
    \begin{equation}\label{asymmetry_two}
\mathbf{S^-_V}(\omega)=-2\hbar\omega {\mathbf G^{ h}_V}(\omega),
\end{equation}
   This generalizes the scalar noise asymmetry obtained in Refs.\cite{lesovik_loosen,gavish}.
 It yields directly the standard equilibrium FDT provided time-reversal symmetry is ensured. In this case the equilibrium noise matrix obeys the detailed balance equation: $
 {\mathbf{S}_{V=0}}(-\omega)=e^{{\beta\omega}}{\mathbf{S}_{V=0}}(\omega).$
 Thus Eqs.(\ref{asymmetry_two},\ref{sa}) yield:
$
{\mathbf{S}_{V=0}}(\omega)=2\hbar\omega N(\omega) {\mathbf G^h_{V=0}}(\omega),
$
where $N(\omega)=1/(-1+e^{\beta\omega})$. Equation (\ref{asymmetry_two}) offers a generalized FDT which extends to both out-of-equilibrium and/or violation of time-reversal symmetry.

It has as well other useful consequences. In case ${\mathbf G^h_V}(\omega)$ is known, Eq.(\ref{symnon}) relates the symmetrized to non-symmetrized current fluctuations.
Similarly, the fluctuations for negative (respectively positive) frequencies can be deduced from those at positive (respectively negative) frequencies. A more interesting alternative is to deduce ${\mathbf G^h_V}(\omega)$ from $\bf S^-(\omega)$, which allows to get rid of any background undesirable noise, being a difference, and is not subject to the limitations on frequencies as the AC conductance. The latter are due to capacitive effects and to the equilibration condition in the reservoirs: $\omega\tau_{in}\ll 1$ where $\tau_{in}$ is the inelastic time, in order to define a quasi-equilibrium distribution \cite{safi,ines_ff_noise,equi}.

Another important feature which Eq.(\ref{asymmetry_two}) clarifies concerns the asymmetry of the excess FF noise, Eq.(\ref{excess}). While many theoretical works used to study the symmetrized noise, it turns out that most experiments are based on quantum detectors measuring the non-symmetrized excess noise \cite{exp_ff,deblock_06}, which has been the subject of few theoretical works with correlation effects \cite{gavish,theo_ff,bena_07,ines_ff_noise}. An important question is under which criteria one can violate the symmetry obtained in a basic coherent conductor \cite{buttiker_revues}, thus giving an evidence for a quantum measurement. It is interesting to discuss the asymmetry of the full excess fluctuation matrix within our formalism, thus to find the criteria for that of excess cross-correlations as well. This can be achieved by using simply Eq.(\ref{asymmetry_two}) (see Eqs.(\ref{sa},\ref{excess}))\cite{note4}:
\begin{eqnarray}\label{excess_asymmetry}
{\mathbf {{\Delta}S}_V}(-\omega)-{\mathbf {\Delta}S_V^{ T}}(\omega)&=&2\hbar \omega{\mathbf \Delta G^h_V}(\omega).
\end{eqnarray}
 The asymmetry between $\Delta S_{nn'}(-\omega)$ and $\Delta S_{n'n}(\omega)$ requires that $\Delta{{G}^h_{nn'}}\neq 0$. This yields a necessary criteria: non-linearity! However this is not sufficient for different terminals $n\neq n'$: one could have $G^h_{nn'}=0$ at any $\bf V$, thus get symmetry of excess cross-correlations even with non-linearity, which clarifies this fact obtained in chiral edges of the FQHE \cite{bena_07}. At the same time, excess auto-correlations were found to be asymmetric in this system, as well as in quantum wires and carbon nanotubes \cite{ines_ff_noise}. Interactions are however not necessary for that. For instance, asymmetry holds in noninteracting systems where the transmission is energy dependent, such as a wire with two barriers \cite{dolcini_07}, hybrid structures \cite{cottet_08}, and Josephson junctions measured experimentally \cite{deblock_06}. Again non-linearity is a common origin for the asymmetry.

A related fact can be shown in the case of a tunneling junction with arbitrary interactions or disorder. For $\omega-qV/\hbar\gg k_BT$, $q$ being the effective charge, we can show that $\Delta S_{nn}(\omega)=0$ \cite{ines_next} while $\Delta S_{nn}(-\omega)=\hbar\omega  \Delta G_{nn}(\omega)$ do not vanish in non-linear systems. This generalizes and explains such behavior obtained in Refs.\cite{bena_07,ines_ff_noise}.

 Considering again any non-linear system, it is interesting to introduce the combination we call the "modified excess noise":
\begin{equation}\label{MEN}
{\mathbf {\breve{\Delta}S}_V}(\omega)={\mathbf{S}_V}(\omega)-2\hbar\omega N(\omega) {\mathbf{G}^h_V}(\omega).
\end{equation}
 It restores symmetry with respect to any positive/negative frequencies:  ${\mathbf{\breve{\Delta}S}_V}(-\omega)={\mathbf{\breve{\Delta}}S_V}(\omega)$.
In view of Eq.(\ref{asymmetry_two}), ${\breve{\Delta}S_V}(\omega)
=(1+N(\omega)){\mathbf{S}_V}(\omega)-N(\omega){\mathbf {S}_V}(-\omega)$. In a linear system, this becomes the excess noise as one can check using the standard equilibrium FDT, and is identical to the combination measured by a detector \cite{lesovik_loosen} having the same temperature as the system.

Now we show how the out-equilibrium FDT, Eq.(\ref{asymmetry_two}), solves the paradox of the negative sign of the excess noise, which looks counter-intuitive as applying a bias is expected to induce more noise, thus the nomination "excess". We focus on auto-correlations as they become real and can be interpreted in terms of emission/absorption spectrum. In two-terminal geometries, they could be negative, such as in Luttinger liquids whether symmetrization is performed \cite{dolcini,recher_06} or not \cite{bena_07,ines_ff_noise}, or without interactions for an energy-dependent transmission \cite{dolcini,cottet_08}. Indeed, for $\hbar\omega\gg k_BT$, the equilibrium noise vanishes thus $\Delta S_{nn}(\omega)=S_{nn}(\omega)$ \cite{gavish}, which can be shown, by a spectral decomposition, to be always positive being the correlator of the same current at terminal $n$. But the absorption excess noise can be negative if $\Delta G_{nn}^h(\omega)$ is negative enough, see Eq.(\ref{excess_asymmetry}). Symmetrized excess noise contains both emission and absorption, thus can be negative too.

To conclude, we have derived a general time-dependent response formula for an arbitrary Hamiltonian depending in a local/non-local and linear/nonlinear way on time-dependent parameters. This yields a microscopic and current-conserving expression of the differential conductance matrix in a multi-probe mesoscopic system with arbitrary time-dependence of the Hamiltonian including the electrochemical potentials in the terminals.  We have deduced a general time-dependent out-of-equilibrium FDT which yields in particular the extension of the equilibrium FDT in case time-reversal symmetry is broken. Its application to the stationary regime has shed light on the asymmetry and sign of the excess FF noise in non-linear systems.

Besides these applications, and that operated in quantum wires \cite{ines_ff_noise}, our theory offers a new promising framework to study systematically time-dependent transport in non-linear systems, including strongly correlated ones, and to consider pumping or mixing setups for instance. Remarkably, we are able to obtain as well higher order differential of any $\langle O(t)\rangle$ with respect to many time-dependent parameters, and therefore that of the current in probe $n_1$ with respect to the potentials in $n_2..n_M$, as will be reported elsewhere \cite{ines_next}.

 The author thanks B. Dou\c cot, P. Joyez and P. Simon for their interest, encouragements and critical readings of the manuscript, as well as M. Polianski, E. Sukhorukov, C. D. Glattli, R. DeBlock, F. Portier and M. B{u}ttiker. She acknowledges
 former collaboration with C. Bena and A. Cr\'epieux and correspondance with A. Cottet.


\begin{thebibliography}{10}
\bibitem{note_modulation}
For a real AC modulation $V_{n'}(t')+v_{n'}(\omega')\cos (\omega'
  t'+\phi_{n'})$, one has $ 2\delta I _{n}(\omega)/\delta v_{n'}(\omega')]_{
  v_{n'}(\omega')=0}=G_{n,n'}(\omega,\omega')e^{i\phi_{n'}}+G_{n,n'}(\omega,-
  \omega')e^{-i\phi_{n'}}$.
\bibitem{equi}
Whenever the reservoirs are of the same nature as the system, as is the case in
  cavities or wires fabricated in two-dimensional gases, one has the condition
  that $L\ll v_F\tau_{in}$ to ensure coherence, thus the equilibration
  condition requires also: $\omega\ll v_F/L$.
\bibitem{note4}
One could as well replace ${\mathbf {\Delta}S_V^{ T}}={\mathbf {\Delta}S_V^{
  *}}$ as it is hermitian.
\end{thebibliography}

\end{document}